\documentstyle[12pt]{article}
\newcommand{\be}[1]{\begin{equation} \label{(#1)}}
\newcommand{\ee}{\end{equation}}
\newcommand{\ba}[1]{\begin{eqnarray} \label{(#1)}}
\newcommand{\ea}{\end{eqnarray}}
\newcommand{\nn}{\nonumber}
\newcommand{\rf}[1]{(\ref{(#1)})}

\def\pmb#1{\setbox0=\hbox{#1}%
  \kern-.015em\copy0\kern-\wd0
  \kern.03em\copy0\kern-\wd0
  \kern-.015em\raise.0233em\box0 }
\def \znbb {0\nu\beta\beta}

\def \emass {\langle m_{\nu} \rangle}
\def \emnu {\langle m_{N}^{(U)} \rangle}
\def \emnv {\langle m_{N}^{(V)} \rangle}
\def \elam {\langle \lambda \rangle}
\def \eeta{\langle \eta \rangle}
\def \exi{\langle \xi \rangle}

\begin{document}
\begin{titlepage}

\begin{center}
{\Large Double beta decay in left-right symmetric models}
\bigskip

{M. Hirsch \footnote{mahirsch@enull.mpi-hd.mpg.de} and 
 H.V. Klapdor-Kleingrothaus \footnote{klapdor@enull.mpi-hd.mpg.de} \\
{\em Max-Planck-Institut f\"ur Kernphysik} \\
{\em P.O. Box 10 39 80, D-69029 Heidelberg, Germany}
}

\bigskip
{O. Panella \footnote{panella@hppg04.pg.infn.it}}\\
{\em Istituto Nazionale di Fisica Nucleare, Sezione di Perugia\\
Via A. Pascoli, I-06123 Perugia, Italy\\
and\\
Laboratoire de Physique Corpusculaire, Coll\`ege de France\\
Place Marcelin Berhtelot, F-75231, Paris Cedex 05, France}

\end{center}

{\begin{center} ABSTRACT \end{center}

{\small \hspace*{0.3cm}
Left-right symmetric models provide a natural framework for 
neutrinoless double beta ($\znbb$) decay. In the analysis of 
$\znbb$ decay in left-right symmetric models, however,  
it is usually assumed that all neutrinos are light. On the other 
hand, heavy {\it right-handed} neutrinos appear quite naturally 
in left-right symmetric models and should therefore not be 
neglected. Assuming the existence of at least one right-handed 
heavy neutrino, absence of $\znbb$ decay of $^{76}$Ge currently 
provides the following limits on the mass and mixing angle of 
right-handed W-bosons:
$m_{W_R}\ge 1.1 $ TeV and $\tan(\zeta) \le 4.7 \times 10^{-3}$
for a particular value of the effective 
right-handed neutrino mass, $\emnv = 1$ TeV, and in the limit of 
infinitly massive doubly charged Higgs ($\Delta^{--}$).
The effects of the inclusion of the  Higgs triplet 
on $\znbb$ decay are also discussed. 
}}

\bigskip
\noindent
{\it PACs:} 11.30, 12.30, 13.10, 13.15, 14.60,14.80, 23.40 

\medskip
\noindent
{\it keywords:} Left-right symmetric models, double beta decay, neutrino 
mass, right-handed W-bosons, Higgs triplet models

\end{titlepage}
\vskip10pt

Left-right symmetric models (LR) \cite{pat74} aim at 
explaining two of the most puzzling questions of the standard 
model (SM), both of which are intimately related to neutrinoless 
double beta decay ($\znbb$) \cite{gk90,doi85}: 
i.) The weak interaction violates parity, and 
ii.) in the standard model neutrino masses are zero. 
Especially if current hints on finite neutrino masses are 
correct, LR models provide a very attractive explanation 
for their small values - when compared to those of the charged 
leptons - via the well-known seesaw mechanism \cite{gel79}.

Of course, $\znbb$ decay has been studied in connection with 
LR models by many theoretical groups before, see 
\cite{hax84} for reviews. However, 
the analysis of $\znbb$ decay is usually either restricted 
to the case where all neutrinos are light 
\cite{doi85,tom86,mut89} or simplified by considering 
only left-handed neutrinos \cite{hax84}. Although 
both approximations look reasonable from a standard model 
point of view, the situation is very different in LR models in 
general. Actually, taking the see-saw mechanism as a motivation 
for LR models one has to expect the existence of some heavy, 
right-handed neutrino. 

The importance of heavy right-handed neutrinos for $\znbb$ decay 
has been pointed out by Mohapatra \cite{moh86}, 
while Doi and Kotani \cite{doi93} derived a quite general 
decay rate, keeping terms for both left- and right-handed 
neutrinos. Both of these papers, however, are not complete. 
While Mohapatra \cite{moh86} considered only the contribution 
proportional to $(m_{W_L}/m_{W_R})^2$, Doi and Kotani 
\cite{doi93} did not calculate the relevant nuclear matrix 
elements. In view of the experimental progress on double 
beta decay \cite{hdmo94,expreview} we therefore felt motivated to 
reconsider $\znbb$ decay in LR models and derive constraints 
on the various parameters of the decay rate in a more general 
way. For this purpose we have calculated matrix elements in 
the limit of heavy neutrino exchange in a pn-QRPA model \cite{mut89,hir94}. 

Furthermore, we discuss modifications of the formalism once the 
contribution of the Higgs triplet is taken into account.
Assuming the 
validity of the SM gauge group and simply adding an Higgs 
triplet to the particle content opens up new decay channels 
for $\znbb$ decay \cite{mv81}, which however were shown to 
be negligible by Schechter and Valle \cite{sch82}, 
Wolfenstein \cite{wol82} and Haxton et al. \cite{hax82}. 
Again, the situation is different in LR models. While an Higgs 
triplet is fairly exotic an extension of the SM, in LR models 
it could provide an attractive explanation of the Majorana 
nature of the neutrino \cite{riz82}. Moreover, Rizzo \cite{riz82} 
has argued that the contribution of the doubly-charged Higgs to 
the {\it inverse $\znbb$ decay} is a necessary ingredient to 
preserve the unitarity of the cross section. Thus, 
although of quite modest numerical importance for limits 
on $W_R$ in usual $\znbb$ decay, 
as we will show at the end of this work, 
we felt the necessity to include the Higgs triplet in 
our analysis. 

As a starting point for the calculation the following effective 
Hamiltonian (in the notation of \cite{doi85}) is used:

\be{ehdoi}
  {\cal H}_W^{CC} = {G\over{\sqrt{2}}}
                \Big\{ J^{\dagger}_{\mu L} j^{-}_{\mu_L} 
                     + \kappa J^{\dagger}_{\mu R} j^{-}_{\mu_L} 
                      + \eta J^{\dagger}_{\mu L} j^{-}_{\mu_R} 
                    + \lambda J^{\dagger}_{\mu R} j^{-}_{\mu_R}\Big\}.
\ee
Here, $J^{\dagger}_{L/R}$ and $j^{-}_{L/R}$ are left- and 
right-handed hadronic and leptonic currents, respectively. 
$\kappa$, $\eta$ and $\lambda$ are the right-handed parameters, 
defined such that the SM charged weak current Hamiltonian 
is obtained in the limit when $\kappa$, $\eta$ and $\lambda$ 
approach zero \cite{doi85}. 

Since $\lambda, \eta \ll 1$ one could think of deriving 
the decay rate considering only contributions of $\lambda$ 
and $\eta$ in lowest order. However, such a procedure leads 
to the neglection of important terms. In general, keeping 
also higher order terms, the decay rate can be written as 
a fourth-order polynomial in $\lambda$ and $\eta$ 
\footnote{Terms proportional to $\kappa$ can be safely neglected 
\cite{doi85}.} as derived in \cite{doi93}. However, to 
separate the particle from the nuclear physics part of the 
calculation, it is convenient to assume that there are no 
neutrinos with mass eigenstates in the range of ${\cal O}$(10-1000) 
[MeV]. Using this well-motivated assumption, after some lengthy 
but straightforward calculation, we write the inverse 
half-life for $\znbb$ decay in a {\it factorized} form as 
\cite{hir95}, 

\ba{fdr1}
\Big[T_{1/2}^{\znbb}(0^+ \rightarrow 0^+) \Big]^{-1} = 
\Big(\frac{\emass}{m_e}\Big)^2 C_{mm}^{LL} \nn
\ea
\ba{fdr2}
+\Big(\frac{m_p}{\emnu}\Big)^2 C_{mm}^{NN}
+ \elam^2 C_{\lambda\lambda}^{LL} 
+ \eeta^2 C_{\eta\eta}^{LL}
+\exi^2 C_{mm}^{NN} \nn
\ea
\ba{fdr3}
+\Big(\frac{\emass}{m_e}\Big)\Big(\frac{m_p}{\emnu}\Big) C_{mm}^{NL}
+\Big(\frac{\emass}{m_e}\Big)\elam C_{m\lambda}^{LL}
+\Big(\frac{\emass}{m_e}\Big)\eeta C_{m\eta}^{LL} \nn
\ea
\ba{fdr4}
+ \Big(\frac{\emass}{m_e}\Big)\exi C_{m\xi}^{NL}
+ \Big(\frac{m_p}{\emnu}\Big) \elam C_{m\lambda}^{NL}
+ \Big(\frac{m_p}{\emnu}\Big) \eeta C_{m\eta}^{NL} \nn
\ea
\ba{fdr}
+ \Big(\frac{m_p}{\emnu}\Big) \exi  C_{mm}^{NN}
+ \elam \eeta C_{\lambda\eta}^{LL}
+ \elam \exi  C_{m\lambda}^{NL}
+ \eeta \exi  C_{m\eta}^{NL}
\ea
where $C_{xy}^{\alpha\beta}$ are products of nuclear matrix 
elements and phase space integrals. In the limit when all neutrinos 
are light, eq. \rf{fdr} reduces to the expression previously 
used \cite{doi85,mut89}. Correspondingly, all coefficients with 
``LL'' superscripts coincide with those of the light neutrino 
case, see \cite{doi85,mut89}. Complete definitions for the 
coefficients for the heavy neutrino case are given in \cite{hir95}.

The particle physics parameters of the decay rate are defined as:

\be{emass}
\emass = {\mathop{\sum_j}}' U_{ej}^2 m_j
\ee
\be{hmass}
\emnu^{-1} = {\mathop{\sum_j}}'' U_{ej}^2 m_j^{-1}
\ee
\be{elam}
\elam = {\mathop{\sum_j}}' U_{ej}V_{ej} \lambda ,
\ee
\be{eeta}
\eeta = {\mathop{\sum_j}}' U_{ej}V_{ej} \eta ,
\ee
\be{exi}
\exi = \Big[\lambda^2 + \eta^2 - 2 \lambda \eta 
            \Big(\frac{M_{GT}^{N}+M_F^N}{M_{GT}^{N}-M_F^N} \Big)\Big]
      {\mathop{\sum_j}}'' V_{ej}^2 \Big(\frac{m_p}{m_j}\Big) .
\ee
$U_{ej}$ and $V_{ej}$ are the elements of the neutrino mixing matrix 
for the left- and right-handed sectors, which satisfy the 
completeness ($\sum_{j} |U_{e j}|^2 = \sum_{j} |V_{e j}|^2 = 1$) 
and the orthogonality relation ($\sum_{j} U_{e j} V_{e j} = 0$) 
\cite{doi85}. As usual the primed sum indicates \cite{doi85} 
that the sums extend over light mass eigenstates only, whereas 
the double primed sums extend over the heavy mass eigenstates.  
$\exi$ describes right-handed neutrino exchange and the 
first term in $\exi$ corresponds to the one considered 
by Mohapatra \cite{moh86}. Neglecting all other terms and assuming 
no mixing between the W-bosons, our decay rate reproduces the 
one considered by Mohapatra \cite{moh86}. Note, that in deriving 
eq. \rf{fdr} we have neglected light right-handed neutrinos, 
as well as terms proportional to 
${\mathop{\sum_j}}'' U_{ej}V_{ej}\lambda m_j^{-2}$ and 
${\mathop{\sum_j}}'' U_{ej}V_{ej}\eta m_j^{-2}$, since the 
latter are suppressed by additional powers of large neutrino masses. 

We have calculated the matrix elements for heavy neutrino 
exchange within the pn-QRPA model of Muto et al. \cite{mut89,hir94} 
Numerical results for the experimentally most interesting 
isotopes are given in table 1. Corresponding matrix elements 
for light neutrino exchange can be found in ref. \cite{mut89}. 
Table 1 shows that, with the possible exception of the two 
heaviest isotopes, all matrix elements have rather similar 
numerical values, in agreement with the expectation.  

\begin{table}[t]
{{\bf Table 1:} {\it Nuclear matrix elements for heavy neutrino exchange in 
$0\nu\beta\beta$ decay for the experimentally most interesting 
isotopes calculated within pn-QRPA.}
}\\
\begin{center}
\begin{tabular}{|c|c|c|c|c|c|c|c|c|}\hline
&&&&&&&&\\
$^{A}$Y & $^{76}$Ge & $^{82}$Se  & $^{100}$Mo & $^{116}$Cd 
& $^{128}$Te & $^{130}$Te  & $^{136}$Xe & $^{150}$Nd \\
&&&&&&&&\\
\hline
&&&&&&&&\\
$M_{GT}^N$ & 236 & 213 & 269 & 156  & 248 & 219 & 121 & 344 \\
&&&&&&&&\\
\hline
&&&&&&&&\\
$M_F^N$ & -53 & -47 & -64 & -36  & -55 & -48 & -27 & -78 \\
&&&&&&&&\\
\hline
\end{tabular}
\end{center}
\end{table}

With the calculated matrix elements at hand, using the half life limit 
on $^{76}$Ge $\znbb$ decay as recently measured by the Heidelberg-Moscow 
collaboration, $T_{1/2}^{\znbb}(^{76}Ge) \ge 7.4 \times 10^{24}$ 
years (90 \% c.l.) \cite{hdmo94}
we are ready to derive quantitative constraints on the various LR 
model parameters. In principle, the experimental half life limit 
and eq. \rf{fdr} 
define an excluded area in a 5-dimensional parameter space. However, 
since in LR models heavy neutrinos are expected to be right-handed, 
we will restrict the discussion to $\emass$, $\elam$, $\eeta$ 
and $\exi$. \footnote{Assuming only left-handed heavy neutrinos 
to contribute to $\znbb$ decay one could also derive the 
constraint $\emnu \ge 6.0 \times 10^7$ GeV. However, $\emnu$ 
incorporates $\Big({\mathop{\sum_j}}'' U_{ej}^2 \Big)^{(-1)}$ 
which has to be expected to be small. Moreover, since the 
{\it effective} masses include the unknown mixing coefficients, 
it has to be noted that $\emnu$ is not necessarily positive 
definite. Thus, by an extreme fine-tuning it is possible to cancel 
the contributions from light and heavy {\it left-handed} 
neutrinos. We disregard such an unlikely situation in the following.} 

Constraints can be derived under the assumption that only one 
parameter contributes to the decay rate at a time (``on axis''), 
or for an arbitrary variation of all four parameters. Numerically 
we find: $\emass = 0.66 (0.56)$ [eV], $\elam =1.1 (1.0) \times 10^{-6}$, 
$\eeta = 6.4 (5.5) \times 10^{-9}$ and 
$\exi = 1.7 (1.7) \times 10^{-8}$ for the ``arbitrary'' (``on axis'') 
cases, respectively. 

As is clear from eq.\rf{elam}-\rf{eeta}, limits on $\elam$ and $\eeta$ 
can not be converted into limits on the mass or mixing 
angle of right-handed W-bosons, without making assumptions 
about the size of the neutrino mixing matrix coefficients 
and their respective $CP$ eigenvalues. Instead, for 
example, $\elam$ defines an excluded area in the 
plane $\{ \mathop{\sum_j}' U_{ej}V_{ej} , m_{W_R} \}$, 
as is shown in fig. 1. Although for large mixing very 
stringent limits would be obtained, for typical values 
of $ \mathop{\sum_j}' U_{ej}V_{ej} \simeq {\cal O}(10^{-6})$ 
or so, only $m_{W_R} \le m_{W_L}$ is excluded, certainly 
not a very stringent constraint.

Much more interesting in this sense is the limit on $\exi$. 
From the completeness relation we know that there is at least 
one right-handed neutrino with 
$V_{ej}^2 \simeq {\cal O}(1)$, which in LR models should be 
quite heavy. Defining 

\be{hmassv}
\emnv^{-1} = {\mathop{\sum_j}}'' V_{ej}^2 m_j^{-1}
\ee
from the limit on $\exi$ one can derive 

\be{limwr}
m_{W_R} \ge 1.1 \Big(\frac{\emnv}{1 \hbox{TeV}}\Big)^{(-1/4)}{\hbox{[TeV]}} ,
\ee
\be{limzeta}
tan(\zeta) \le 4.7 \times 10^{-3} 
            \Big(\frac{\emnv}{1 \hbox{TeV}}\Big)^{(1/2)}.
\ee
To compare the limit on the mass of the $W_R$ to the one derived 
by Mohapatra \cite{moh86}, we mention that 
$m_{W_R} \ge 1.23 \Big(\frac{\emnv}{1 TeV}\Big)^{(-1/4)}{\hbox{[TeV]}}$ 
can be derived, if we take the limit $tan(\zeta)\rightarrow 0$. 
Finally, following the argument \cite{moh86} that vacuum stability 
requires $\emnv \le g m_{W_R}$, where $g$ is of order 
${\cal O}(1)$, an absolute lower bound on $m_{W_R}$ 
can be derived. The 
quantitative difference between our result and that of ref. 
\cite{moh86} is mainly due to the improved half life limit used 
in our calculation. We stress that it is {\it not} due to 
errors or uncertainties in the matrix element calculation - 
uncertainties of limits on $m_{W_R}$ scale only as the fourth 
square root of the uncertainties in the nuclear matrix 
elements. 

Let us now turn to a brief discussion of the Higgs triplet contribution 
to $\znbb$-decay. 
The generation of  Majorana masses in left-right symmetric models
is achieved quite naturally if the Higgs sector of 
the theory contains 
two additional triplets, 
$\Delta_{L/R} = (\Delta^{--},\Delta^{-},\Delta^{0})_{L/R}$\cite{riz82a}. 
This implies that $\znbb$ decay can not only occur through the 
usual neutrino exchange diagram (fig. 2.a), but in addition also through the 
graph involving the exchange of a doubly-charged 
Higgs (fig. 2.b).~\footnote{In addition, there is the possibility 
that the two W-bosons 
of fig. 2.a are replaced by the singly-charged component of the 
triplet. This contribution, however, is negligible due to the 
small coupling of the Higgs to quarks \cite{sch82}, and, in 
addition, due to the small relevant nuclear matrix elements 
\cite{hax82}.} It is straightforward to show that the contribution 
of this graph to $\znbb$ decay is proportional to 
$\frac{1}{m_{W_R}^4}\frac{m_N}{m_{\Delta^{--}_R}^2}$ \cite{riz82a}. 
(In general, also $\Delta_R$ and $\Delta_L$ can mix with each 
other as is the case for the W-bosons. The left-handed doubly-charged 
Higgs, however, has  a negligible coupling strength
proportional to the light 
neutrino mass. We neglect this possibility for simplicity.) 

The inclusion of the graph in fig 3.b therefore modifies the 
contribution of the $\lambda^4$-terms, which are then 
proportional to, (neglecting mixing among neutrinos for 
simplicity)

\be{modif}
\Big(\frac{m_{W_L}}{m_{W_R}}\Big)^4 
\Big( \frac{1}{m_N} + \frac{m_N}{m_{\Delta^{--}_R}^2} \Big).
\ee

Eq. \rf{modif} leads to a modified constraint on the mass of the
right-handed W-bosons as shown in fig. 3, where limits are 
shown as a function of the heavy, right-handed neutrino mass and 
various values of $m_{\Delta^{--}_R}$. Given that there is 
no {\it upper} bound on the mass of the right-handed Higgs triplet, 
however, no more stringent constraints on $m_{W_R}$ can be inferred 
from $\znbb$ decay, than the one quoted in eq. \rf{limwr}. 

To summarize, it is concluded that right-handed 
neutrino exchange in $\znbb$ decay leads to much more 
stringent limits on the mass and mixing angle of right-handed 
W-bosons, than the left-right mixing mechanism, usually 
expressed in terms of the {\it effective parameters} $\elam$ 
and $\eeta$. This is mainly 
due to the small mixing, which has to be expected between 
the left- and the right-handed neutrino sectors. 
Terms proportional to right-handed neutrinos can not 
be neglected in the decay rate of $\znbb$ decay in 
left-right symmetric models. Although limits on the mass 
of the right-handed W-boson do depend only weakly on nuclear 
matrix elements, it therefore seems to be desirable also to 
reconsider the calculation of nuclear matrix elements for heavy 
particle exchange more carefully than has been done up to now. 

We have also discussed the modifications of the formalism, 
due to the contribution  
of the right-handed Higgs triplet. 
Although 
such contribution turns out to be numerically small
(unless $\Delta^{--}$ is very light) 
for $\znbb$ decay, as is shown in fig. 3, it is
necessary to include these 
terms if one wants to make a consistent comparison of the 
constraints on LR models as derived from $\znbb$ decay with 
those inferred from inverse neutrinoless double beta decay 
searched for at particle accelerators \cite{bel95,riz94}. 

\bigskip
\begin{center}
ACKNOWLEDGMENTS
\end{center}
We would like to thank S.G.Kovalenko for useful discussions. 
M.H. is supported by the Deutsche Forschungsgemeinschaft 
(kl 253/8-1 and 446 JAP-113/101/0). O.P. acknowledges partial support 
from the E.U. program ``Human Capital and Mobility'' under contract No.
CHRX-CT92-0026. He would also like to thank the Max--Planck--Institut
f\"ur Kernphysik in Heidelberg for the very kind hospitality.

\bibliographystyle{unsrt}

\medskip

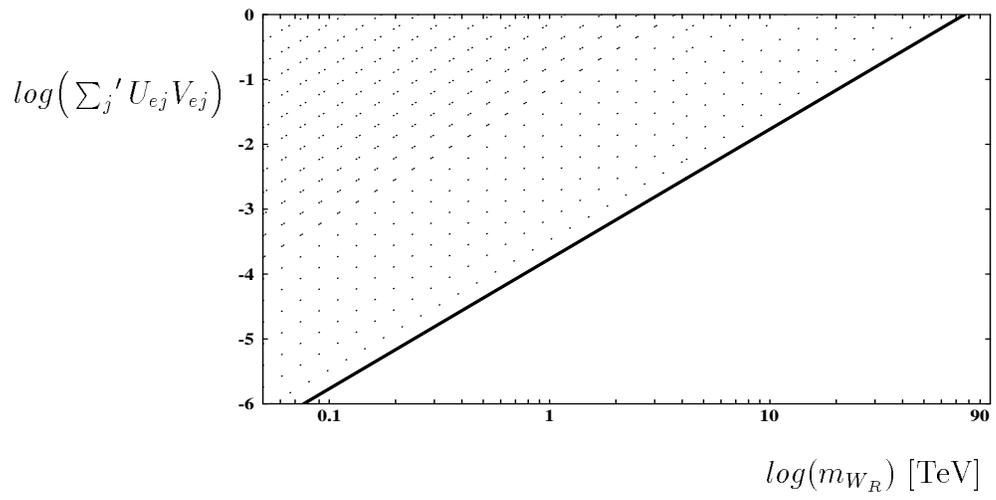
{\bf Fig. 1: }{\it Excluded area in the plane 
$\{ \mathop{\sum_j}' U_{ej}V_{ej} , m_{W_R} \}$, for $^{76}$Ge. 
Combinations to the upper left of the thick line are not 
allowed.}


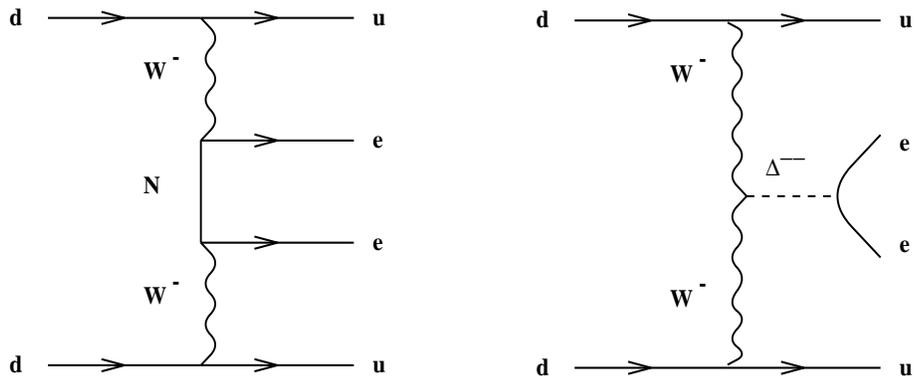
{\bf Fig. 2: }{\it a) To the left: Heavy neutrino exchange contribution 
to neutrinoless double beta decay in left right symmetric models, and 
b) to the right: Feynman graph for the virtual exchange of a 
doubly-charged Higgs boson, see text.}


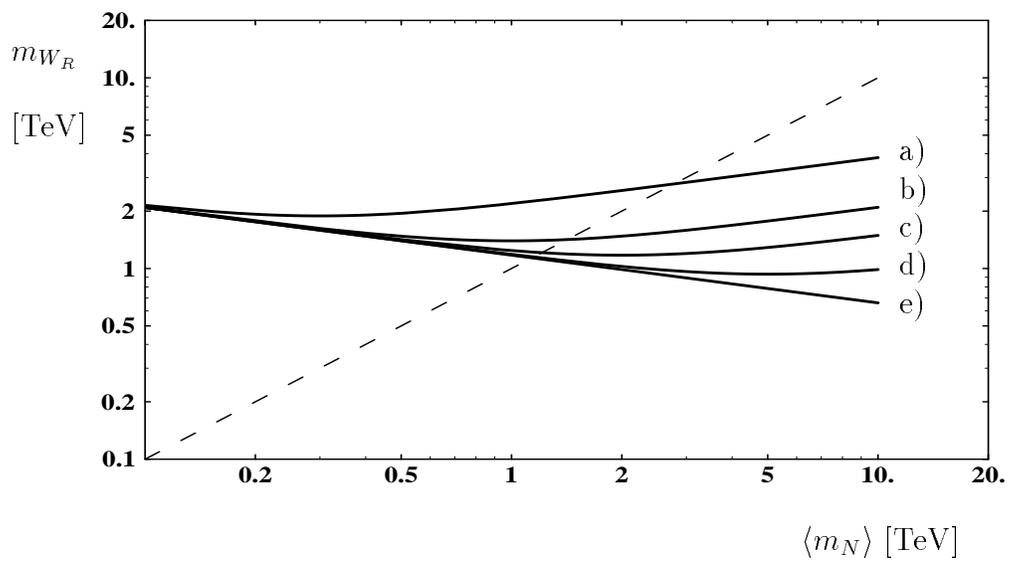
{\bf Fig. 3:} {\it Limits on the mass of the right-handed W-boson from 
neutrinoless double beta decay (full lines) and vacuum stability 
(dashed line). Combinations below the lines are forbidden. 
The five full lines correspond to the following 
masses of the doubly charged Higgs, $m_{\Delta^{--}}$: a) 0.3, 
b) 1.0, c) 2.0, d) 5.0 and e) $\infty$ [TeV]. }

\end{document}